\documentstyle[preprint,12pt]{aastex}
\newcommand{\ngc}{NGC~6752}
\begin{document}
\title{Central Proper-Motion Kinematics of \objectname[]{NGC
6752}\footnote{ Based on observations made with the NASA/ESA Hubble
Space Telescope, obtained from the data archive at the Space Telescope
Science Institute. STScI is operated by the Association of
Universities for Research in Astronomy, Inc. under NASA contract NAS
5-26555.}}

\author{G.A. Drukier, C.D. Bailyn, W.F. Van Altena, T.M. Girard}
\affil{Dept. of Astronomy, Yale University, P.O. Box 208101, New Haven, CT, 06520-8101, USA}
\email{drukier, bailyn, vanalten, girard@astro.yale.edu}

\begin{abstract}
We present proper motions derived from WFPC2 imaging for stars in the
core of the peculiar globular cluster \ngc. The central velocity
dispersion in both components of the proper motion is 12 km
s$^{-1}$. We discuss the implications of this result as well as the
intrinsic difficulties in making such measurements. We also give an
alternative correction for the 34-row problem in the WFPC2 CCDs.
\end{abstract}

\section{Introduction}
With HST now having been in orbit for over 10 years, we can now
combine high resolution images over a sufficiently long baseline to
measure proper motions in the nearer globular clusters. Proper motions
give us two further components to each star's 6-dimensional phase
space coordinate. When combined with radial velocities and positions,
five-sixths of the phase space can be mapped out, leading to greater
understanding of the structure and evolution of globular
clusters. Radial velocities have been measured for many globular
clusters, although the number where there are sufficient measurements
for good statistics is still small. This is, in part, a consequence of
the amount of telescope time required to acquire spectra of many
stars.  Fiber-fed spectrographs such as HYDRA or 2DF go some way to
alleviating this problem, but fiber crowding restricts their use in
the centers of globular clusters. Fabry-Perot techniques use direct
imaging to measure radial velocities in more crowded regions, but
their spectral resolution and range is limited, and calibration can be
difficult. Proper motions, on the other hand, rely only on direct
imaging, and fairly short exposures suffice. The time we are required
to wait between exposures to measure a given velocity increases with
the distance of the object, but decreases with the telescope's spatial
resolution.

At about 4 kpc, \ngc\ is a good candidate for a proper-motion
study. It is a prominent southern cluster, which has revealed some
unusual features in recent years. It is not clear whether \ngc\ is truly
anomalous in some way, or whether the wealth of observational detail
available has uncovered features that are present in many other less
well-studied clusters as well.

Exposures of \ngc\ taken with the HST pointed at the position of the
cluster center according to \citet{sw86} clearly miss the highest
density region by about 10\arcsec. Examination of Figure 1 of
\citet{rb97} clearly demonstrates this. The method Shawl \& White used
is easily misled by bright foreground stars and this appears to be the
case here. A bright star some 101\arcsec\ north-west of the cluster
may have biased the determination of the center of the cluster (R.
White, 2002, private communication). Instead, we use the center
determined by \citet{rb97}, which is given only in their pixel
coordinates. We have used the World Coordinate System information in
the HST frame header to get J2000 equatorial coordinates of
$\alpha=19^h 10^m 52\fs 16,\delta = -59\degr 59\arcmin 4\farcs 0$ This
is probably good to better than $0\farcs 2$.

\citet{dk} identify \ngc\ as a ``post-core-collapse'' globular cluster
on the basis of its density profile, which does not fit a King model.
However the profile is also poorly fit by a power law.  Furthermore,
\citet{rb97} have identified a high binary fraction in the core, and
dynamical models suggest that a true core-collapse phase cannot be
achieved until most of the binaries have been ``burned''
\citep{gao}. Interestingly, the binary fraction appears to drop
sharply beyond 11\arcsec\ from the cluster center \citep{rb97}.  The
drop-off is much sharper than would be expected from stars of twice
the turnoff mass.

 \ngc\ is among the clusters showing a bimodality in CN abundances on
the giant branch \citep{bimodal}. It was recently found that in
47~Tuc, this bimodality exists on the main sequence itself
\citep{cannon,harbeck}.  There is some evidence that this is also the
case in \ngc\ \citep{gratton}.  The horizontal branch of \ngc\ is blue
as befits its metallicity ([Fe/H]$=-1.64$), but its extent is
unusual. This is one of the clusters where the horizontal branch
features an extreme blue tail reaching stars with $T_{eff}\sim 32,000$
K \citep{mom02}.  The kinematics of the cluster itself are also
unusual. \citet{dinescu} find it to be the oldest, most metal-poor
cluster with a measured space velocity to have disk-like
kinematics. In its kinematics, it is quite similar to 47 Tuc despite
being much more metal poor. That an old, metal-poor cluster can have
the kinematics of the thick disk has interesting implications for the
history of the Galaxy.

A variety of interesting compact objects has been reported in \ngc\ .
A recent Chandra X-ray study \citep{pooley} has identified 19 sources
within the half mass radius of the cluster, of which over half are
apparently CVs belonging to the cluster.  This contrasts with the
results from 47~Tuc \citep{grindlay} in which most of the faint X-ray
sources appear to be millisecond pulsars.  However, radio observations
have identified five millisecond pulsars in \ngc\ \citep{damico1}.
Three of these are close to the center of the cluster, and two of
these three have negative period derivatives. These set lower limits
on the gravitational acceleration of the pulsars by the cluster
potential. D'Amico et al. set a lower limit of 10 on the mass-to-light
ratio in solar units inside 7\arcsec. This, however, was based on the
\citet{sw86} center. More recently, they have also recognized the
deviation of the true center from that of \citet{sw86}. Using a center
very close to ours, the pulsars lie closer to the center and the new
lower-limit on the mass-to-light ratio is 6 in solar units
(A. Possenti, 2002, private communication).  This is still greater
than what would be suggested by the only previous velocity study of
\ngc\, in which integrated-light spectra by \citet*{dmm97} yielded a
central velocity dispersion of 4.5 km s$^{-1}$. As we will see below,
we obtain a significantly higher velocity dispersion, so this may not
be a problem.

The outermost of the millisecond pulsars is in a $P_{\rm orb}=0.86$
day circular binary at least 3.3 half-mass radii (8 pc) away from the
cluster center \citep{damico1}. This is an unusual place to find a
binary millisecond pulsar, and the circular orbit sets firm limits on
how it came to be there. \citet{colpi} argue that the most likely
scenario leading to this system's present location is one where the
pulsar system is scattered by a binary black hole with total mass of 3
to 100 $M_\sun$. Such a binary black hole is probably distinct from
any black hole which may form part or all of the dark component
implied by the high central mass-to-light ratio.

As noted above, it is not clear whether these various anomalies mean
that \ngc\ is unusual, or simply well-studied.  But either way, a more
detailed understanding of its dynamics would appear to be important.
The remainder of this paper is in four parts. In the next section we
discuss in detail our observations and reduction procedure. Following
that we present our velocity dispersion profiles. The methods for
inferring such kinematic conclusions from the data are, in our
opinion, unsatisfactory. We discuss this in \S\ref{discussion}. We sum
up our results in the concluding section.

\section{Observations and Reduction}
\label{obs}
Our proper motions are based on two epochs of WFPC2 data from HST.  The
earlier epoch consists of 116 PC-only frames taken as part of program
GO~5318 on 1994 August 18-19. These were part of a search for variable
stars in the center of the cluster.  In this study we have used the
26.0 s F555W frames. These are arranged in a $3\times 3$ dither pattern
centered on the \citet{sw86} center. We shall refer to this as the
``E1994'' data.

The second epoch consists of 12 WFPC2 frames taken as part of program
GO~7469 on 1999 September 12. There are 4 frames at each of 3
positions arranged in an `L'-shaped pattern with offsets between them
the size of a PC field.  The original intention had been to reproduce
the position of frames taken with the original PC in 1991 August as
part of program GTO~2943. In these data the center of the cluster
appears principally on the WF chips. This is due to the error in the
assumed center, as well as a measure of confusion in defining the
offsets required to put the WFPC2 PC in the same places as in the 1991
data. We shall refer to this epoch as ``E1999''. 

The result of the combination of these two data sets is that while the
area is limited to that of one PC frame by the E1994 data, our proper
motions will be limited by the measurement precision of the E1999
positions.

All the frames were reduced by the methods of \citet{ak00}. This uses
the images of stars observed in multiple frames to measure the
effective point-spread-function in the frame. This ePSF takes into
account the effects of the pixellation on the undersampled WFPC2
images and allows us to measure the position of point sources to a
high precision. Our reductions leave us with 116 raw star lists for
the 1 field covered by the E1994 data and 48 lists for the
over-lapping fields of the E1999 data. What is now required is to
correct the positions in these lists for two things: 34-row error and
distortion in the optics.  The 34-row error \citep{ak99} was corrected
using the model described in the Appendix. Distortion in each WFPC2
chip was corrected using the results of \citet{ak02}.

For the E1994 data, the corrected positions were transferred into a
reference frame by polynomial transformations including terms up to
second order in the positions. The quadratic terms were required to
allow for residual variation in the distortion terms. The reference
frame consisted of the weighted means of the translated
positions. These were stable after two iterations. An iterative
$3\sigma$ clipping algorithm was employed in producing the mean
positions. Any observation that deviated by more than $3\sigma$ in
either dimension was rejected in producing the mean position.

The combination of the E1999 data was complicated by the multiplicity
of individual pointings of the various frames. What we ultimately did
was to first rigidly transform the corrected positions for each frame
to a common one. The data from each CCD was transformed
independently. The mean positions in this common frame formed the
initial approximation to a reference frame. The rigidly transformed
lists were then iteratively transformed onto the reference frame and
then combined as discussed in the previous paragraph to form a new
reference frame. Several iterations were required before the
difference in positions between two successive approximations were
reduced to insignificance. Both quadratic and cubic polynomials were
used for the final transformations, but the final positions agreed
within the estimated errors. We also tried local transformations using
the 100 nearest neighbors to each star, but, again, there was no
improvement in the quality of the final positions.

The final positions in each epoch were calculated as weighted means of
the translated positions in each frame.  The uncertainty in the
positions were estimated from the standard deviation of the individual
positions after the rejection of $3\sigma$ outliers.

To measure dispersions, it is important to have confidence in our
estimate of the errors in the individual positions.  In order to
validate our error estimate, we divided the E1999 data into two halves
and calculated the positions and errors for each half separately. (Our
error budget for the proper motions is dominated by the uncertainty in
the E1999 positions. Since the E1994 data consists of 116 PC frames,
the uncertainties there are an order-of-magnitude smaller.) For each
star we calculated the position differences in each dimension divided
by the standard deviation in the difference. If the errors are correct
and normal, these should be distributed as a Gaussian with zero mean
and unit standard deviation. What we found was that while the mean was
consistent with zero, the standard deviation was 0.90, indicating the
errors to be underestimated. All errors in the final E1999 data were
then divided by 0.90 to correct for this. As will be seen, the
measured velocities are large enough that this 10\% effect in the
error estimation will have no appreciable effect on the final velocity
dispersions.

The final E1994 and E1999 star lists were then matched and a global
quadratic transformation was calculated between them using the E1994
positions as the reference. Cubic and local transformations give the
same results with larger errors for the local transformations.  The
differences between the transformed positions were then taken as the
proper motions in the E1994 coordinate system. A plate scale of
0.04558\arcsec\ pixel$^{-1}$ appropriate to the PC1 chip \citep{wfpc
manual} and the epoch interval of 5.07 years were used to convert the
proper motion into units of mas yr$^{-1}$. The proper motions in each
dimension were examined for systematic effects with respect to the
stellar coordinates in both epochs. None were seen. We also calculated
two proper motions for each star by dividing the frames at each epoch
into two samples. The proper motions in each half-sample were
consistent within the estimated errors. Thus, we are confident that
these have been estimated correctly.

\begin{figure}[t]
\epsscale{0.7}
\plotone{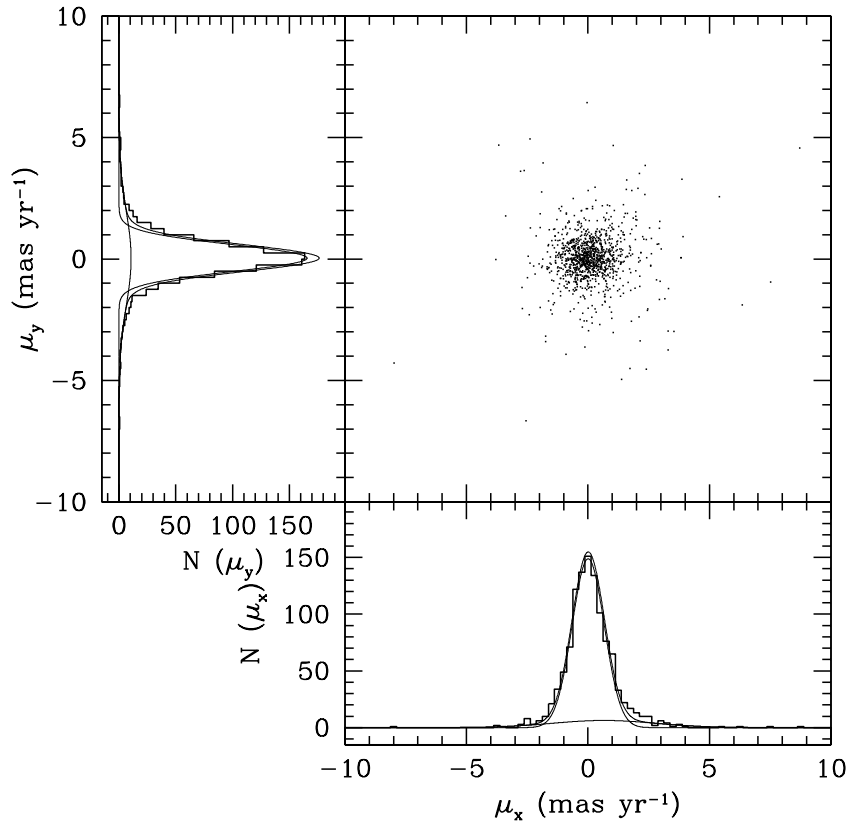}
\caption{We show the $x$ and $y$ components of the proper motions in
the E1994 coordinate system. No field population is apparent in this
diagram.  The histograms highlight the distributions as a function of
proper motion in each component. The fitted two-component Gaussian
model has been superimposed on the histograms. Note the tight central
core and outlier populations.  \label{F:xyhistogram}}
\end{figure}

In total there are 19888 stars with four or more observations in the
E1999 data.  We require this number of observations in order to be
able to make reliable error estimates. Our experience when dividing
the sample into halves suggests that for these data four observations
should be sufficient to estimate the errors adequately. 1281 of these
stars match stars included in the E1994 sample.  For these stars the
median combined error of the two proper motion components is 0.31 mas
yr$^{-1}$, with a mode of 0.17 mas yr$^{-1}$. The distribution of
uncertainties drops off rapidly, so we have established a cut-off at 1
mas yr$^{-1}$ and retained all 1121 stars with errors less than
this. It should be noted that the stars rejected are {\it not}
exclusively those with large proper motions.  The proper motions of
the retained stars are shown in the upper-right of
Figure~\ref{F:xyhistogram}. Note that no field population is apparent
in this diagram. This is not unexpected given the relatively high
galactic latitude of \ngc. There is a similar lack of non-cluster
stars in the color-magnitude diagram of \citet{rb97}.

Also in Figure~\ref{F:xyhistogram} we show histograms of the $x$ and
$y$ components of the proper motions in E1994 coordinates.  Analysis
of the distributions show that they are not described well by single
Gaussians. In each component there is a central core with the addition
of a second, more broadly distributed component. We interpret the
latter as likely due to blended stellar images in the E1999 WF data
being matched with single stars in the E1994 data and giving
spuriously high proper motions. We fit a two-Gaussian mixture model
\citep{nemec2} to the proper motions in each component. Then, for each
star, we calculated the probability that the star was in each
component and a joint probability that the star was in the narrow
component. The mixture models had 12 to 18\% of the stars in the broad
component and, in the joint distribution, 15\% had probabilities of
being in the narrow component of less than 50\%.

\begin{figure}[t]
\plotone{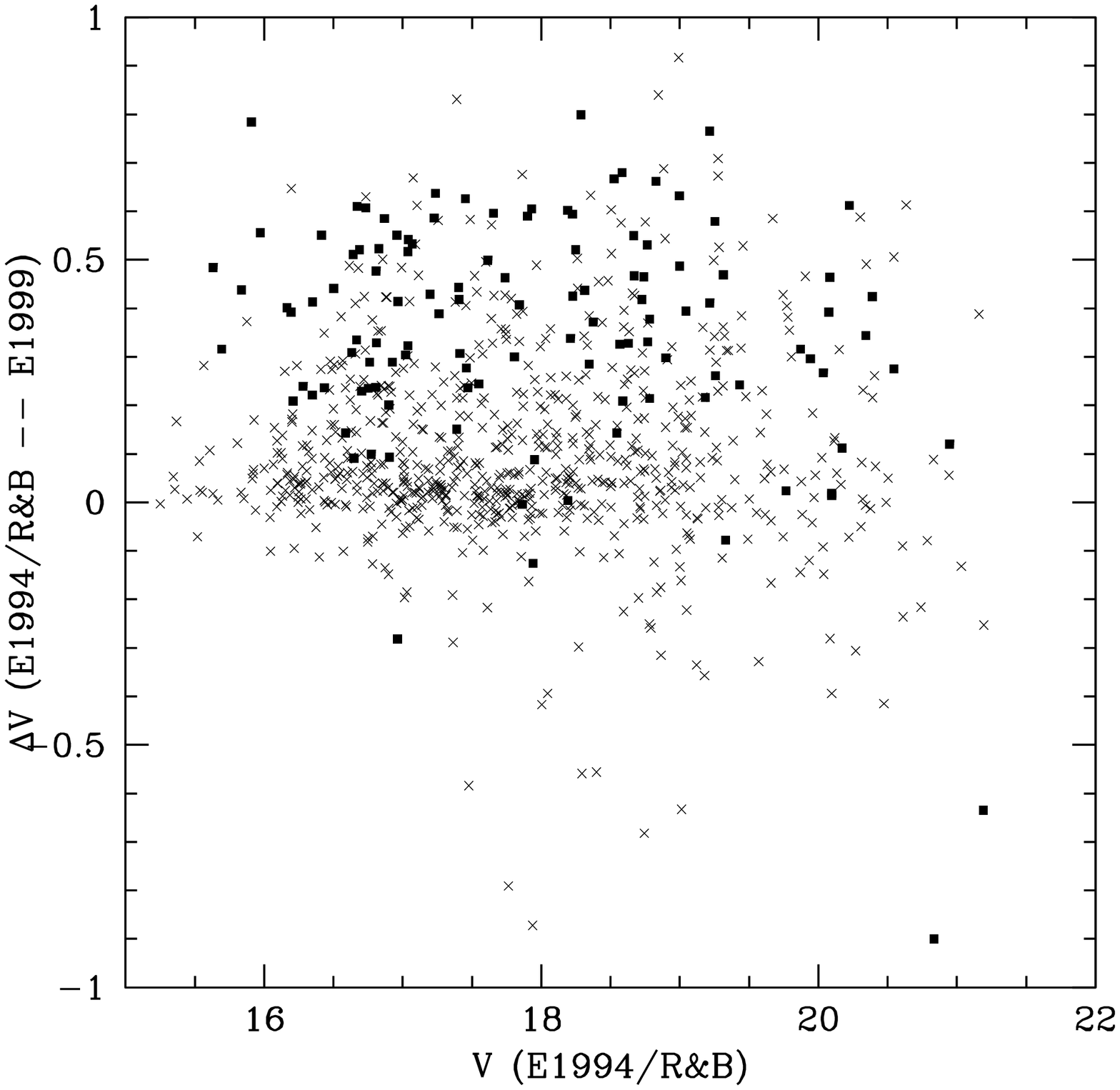}
\caption{Magnitude difference between E1994 and E1999 for all stars
with photometry from \citet{rb97}. Stars with low probability of
membership in the narrow population are shown as squares. These stars
tend to be brighter in the lower resolution E1999 WF
data. \label{F:blended}}
\end{figure}

As support for this interpretation we show in Figure~\ref{F:blended}
the differences between the measured $V$ magnitudes in the E1999 and
E1994 data sets. For the latter we have used the photometry of
\citet{rb97}. The stars with joint probabilities of less than 50\% are
shown as solid symbols.  These are found preferentially on the bright
side of the distribution. This is as we would expect if the images in
the E1999 data were the combination of two stellar images. The
low-probability stars are concentrated to the center as we would
expect. A KS test gives a probability of zero that the two groups come
from the same radial distribution. Even in the most crowded regions
the majority of the stars are well measured, however. Only 19\% of the
stars within 11\arcsec of the center have low probabilities.  This
rises to 26\% within 5\arcsec and 37\% for the 35 stars within
3\arcsec of the center.

\begin{figure}[t]
\plotone{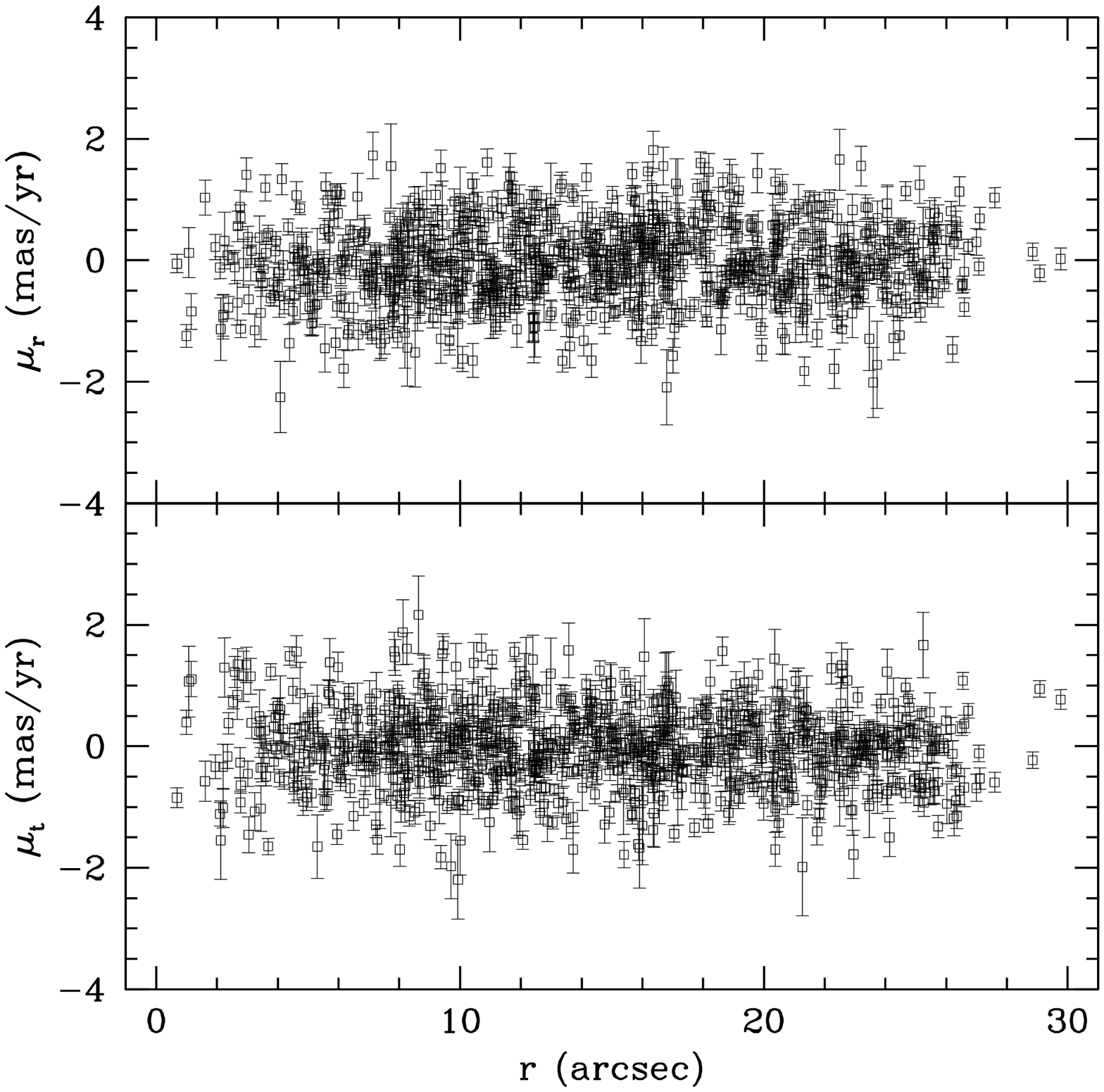}
\caption{The radial (top) and tangential (bottom) components of the
proper motion plotted against distance from the cluster
center.\label{F:proper motions}}
\end{figure}

\begin{figure}[t]
\plotone{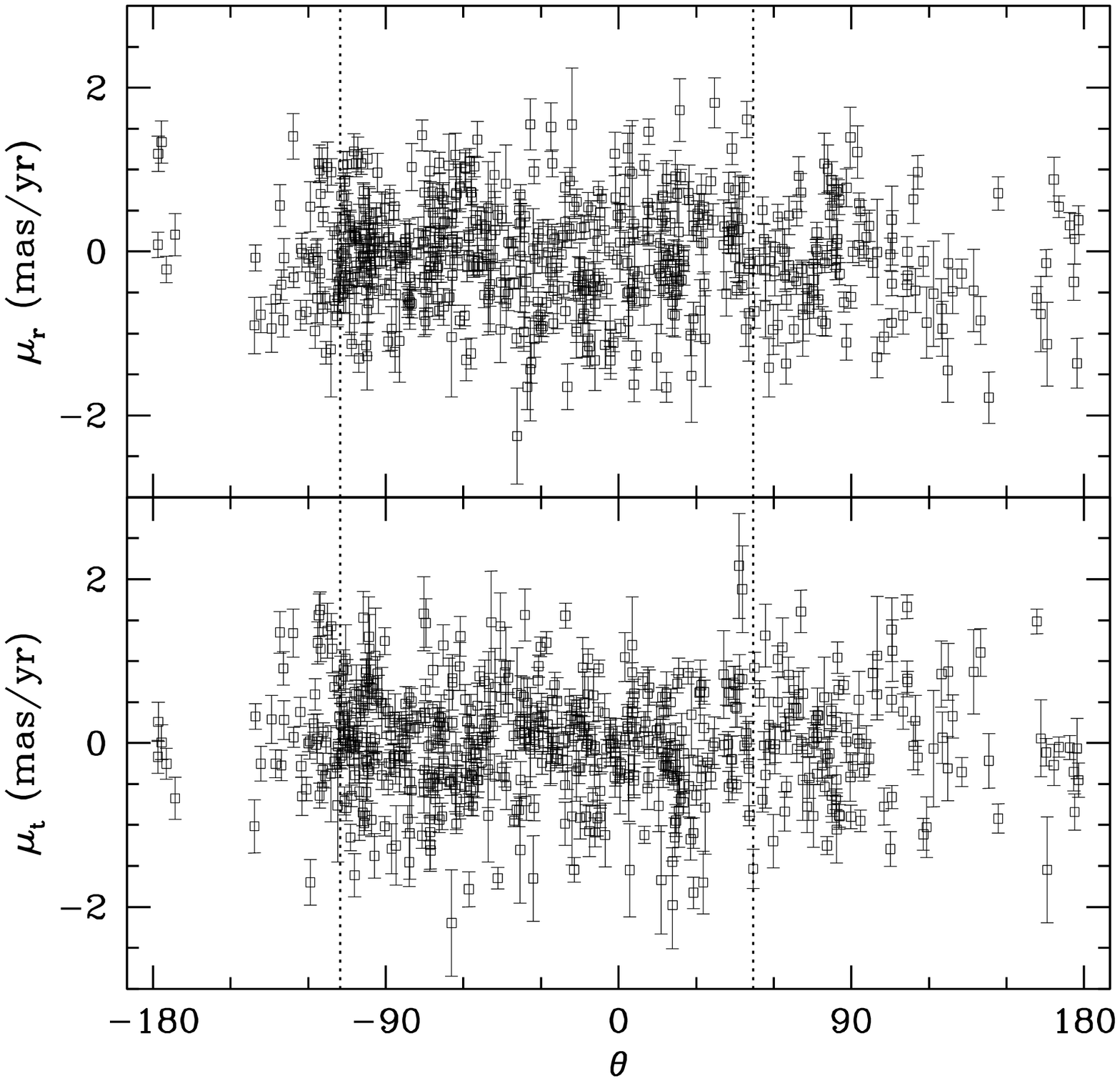}
\caption{The radial (top) and tangential (bottom) components of the
proper motion plotted against angular position about the cluster
center for stars within 16\farcs 7 of the cluster center.  Only the
stars between the vertical lines have been used in measuring the
velocity dispersion. Outside this azimuthal region the radial extent
of the data is truncated by the edge of the E1994
frame.\label{F:azimuthal mu}}
\end{figure}

Rejecting those stars with joint probabilities less than 50\%, we have
950 stars left.  Our final step was to resolve the proper motions into
radial and tangential components with respect to the radius vector
between each star and the \citet{rb97} center. The proper motions as a
function of radius and azimuth are presented in Figures~\ref{F:proper
motions} and \ref{F:azimuthal mu}. The azimuthal sample we use below
is a 431 star subsample of the full star list. They have been selected
to cover a uniform radial region so as not to introduce any biases
arising from any radial variations in the velocity dispersion. The
stars selected cover $160\degr$ about the cluster center to $16\farcs
7$. The E1994 frames lie within 4\arcdeg of North along the $x$ and
East along the $y$ axes, so the azimuths in Figure~\ref{F:azimuthal
mu} run roughly from North through East.

\section{Velocity Dispersion Profile}
Extracting kinematic information from a set of velocities is as
traditional as it is fraught with difficulty. It is customary to
interpret the observed velocities as coming from a Gaussian
distribution and use them to measure the dispersion in the
distribution, usually as a function of position. There are certainly
reasons to believe that this is not actually the case, but, in the
absence of sufficient data to attempt to recover the distribution
function non-parametrically (as in, for example \citealt{merritt w
cen}), or the use of models transferred into the observational domain,
it suffices as a first cut at interpreting the velocity data. It is in
this spirit we calculate the velocity dispersions for \ngc\ from our
proper motions.

In order to investigate the radial and azimuthal dependence of the
velocity dispersion we have binned the stars into equal number bins
and used a maximum-likelihood technique to estimate the dispersion in
each bin. For a given star with observed velocity $v_i$ and
uncertainty $\epsilon_i$, if the systemic velocity is $\bar v$ and the
velocity dispersion in its bin is $\sigma_r$, for bins
$r=1,...,N_{b}$, then the likelihood of measuring $v_i$ is given by
\begin{equation}
L(v_i|\bar v, \sigma_r, \epsilon_i) =
{1\over{\sqrt{2\pi\left(\epsilon_i^2+\sigma_r^2\right)}}}
\exp\left[-{{\left(v_i-\bar v\right)^2}\over{2\left(\epsilon_i^2+\sigma_r^2\right)}}\right].
\label{E:like1}
\end{equation}
For each bin, the total likelihood is the product of eq.~\ref{E:like1}
for all stars in the bin. From the data $\bar v$ is consistent with
zero and we have assumed this value in what follows.  For the radial
sample we have used 5 bins of 190 stars each. For the somewhat smaller
azimuthal sample we have used 3 bins of 108 stars and one of 107.  The
$1\sigma$ uncertainty in the dispersion estimate has been taken as the
half-width of the symmetric region about the maximum likelihood
containing 68\% of the probability. For the binning used, this
uncertainty in the measured dispersion is approximately 4\% for the
radial bins and 6\% for the azimuthal ones.

\begin{figure}[t]
\plotone{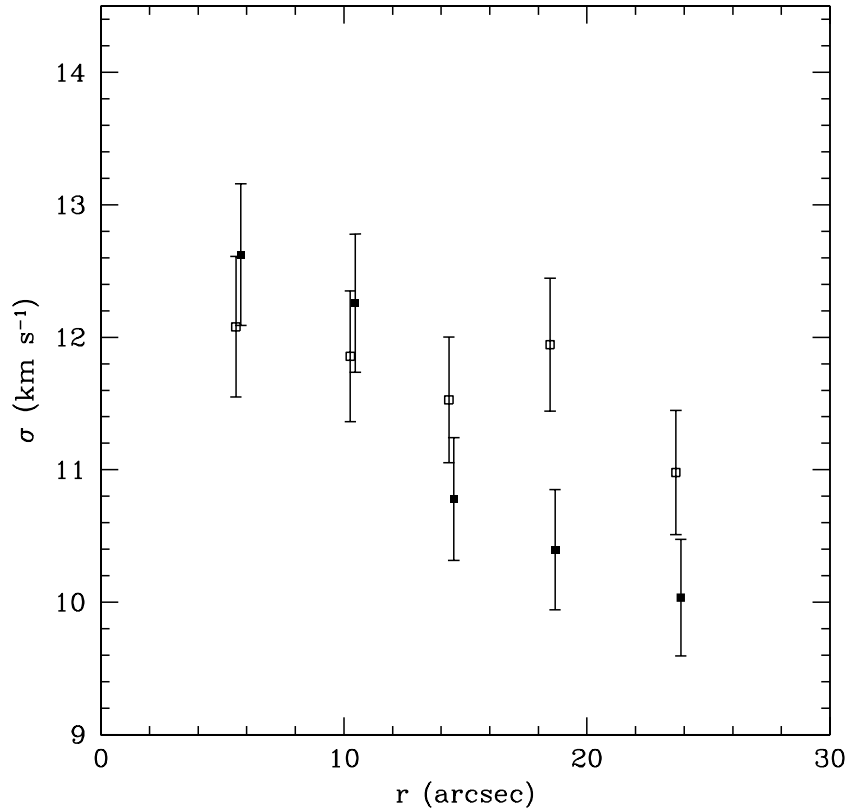}
\caption{The maximum-likelihood velocity-dispersion-profile as a
function of radius from the binned proper motion data. Solid points
are for the tangential component and open points are for the radial
component. For clarity, the radial positions of the points have been
offset $\pm 0\farcs1$  from the mean radius in each
bin.\label{F:radial profile}}
\end{figure}

Figure~\ref{F:radial profile} shows the velocity dispersion profile in
the radial direction for the two components. The proper motions have
been converted into velocities assuming a distance of 4.17 kpc. This
corresponds to a distance modulus $(m-M)_0 = 13.1 \pm 0.1$, and
represents a compromise between \citet{renzini} who found a white
dwarf distance modulus of $13.05 \pm 0.1$  and
\citet{carreta} who find $(m-M)_0 = 13.21 \pm 0.04$ using Hipparcos
subdwarf distances. At this distance 1 mas/yr = 19.77 km s$^{-1}$.

Inside 11\arcsec, i.e. the central, binary-dominated core, the
dispersions in the two components are consistent and flat with a
dispersion of about 12.4 $\pm 0.5$ km s$^{-1}$. This is over twice the
dispersion in the radial velocity as measured by \citet{dmm97} and is
inconsistent with their value. If we adopt the surface brightness
profile parameters of \citet{tkd}, then our velocity dispersion
suggests a central $V$-band mass-to-light ratio of 9 in solar
units. This is consistent with the lower limit from the pulsar
timings, but much larger than that typical for galactic globular
clusters \citep{mclau}.

Outside the core, the dispersion drops  with radius and it appears
that the dispersion in the tangential component drops more quickly
than does the dispersion in the radial component. The presence of such
radial anisotropy so close to the core would be quite unexpected. In the
sophisticated N-body models of \citet{bm02}, for example, little
anisotropy at all is seen in the inner half of the models, and what
there is is tangentially anisotropic, not radially as is the case
here. The differences between the radial and tangential dispersions
are small, at most $2\sigma$, but there is a systematic trend that is
highly suggestive.

\begin{figure}[t]
\plotone{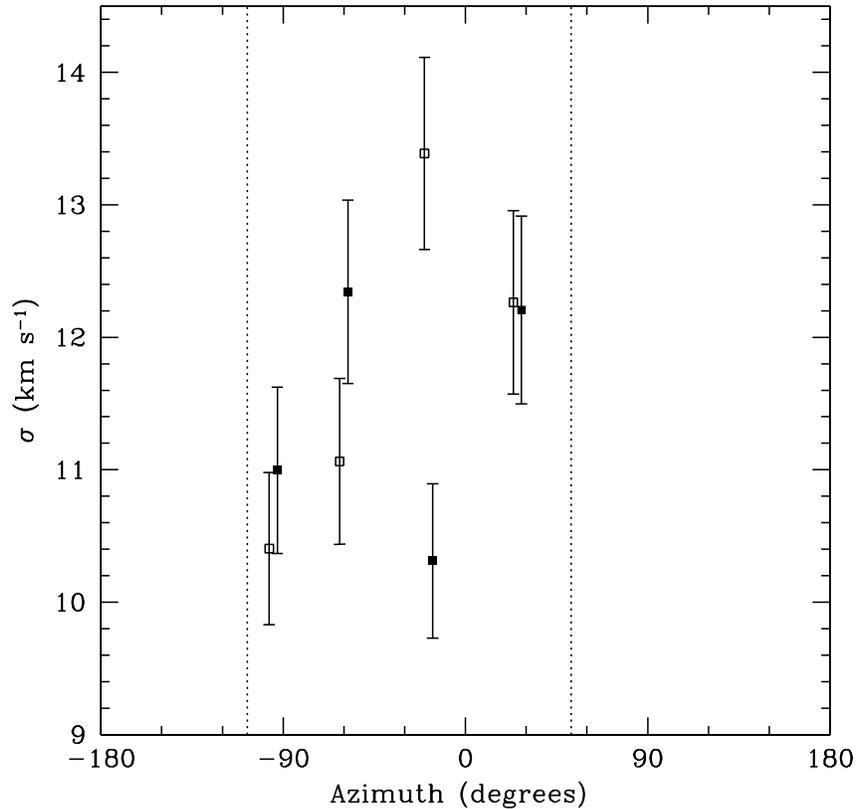}
\caption{Velocity dispersion profile as a function of azimuth about
the cluster center from the binned proper motion data. Solid points
are for the tangential component and open points are for the radial
component. For clarity, The positions of the points have been offset
$\pm 2\degr$ from the mean angle in each bin.\label{F:azimuthal
profile}}
\end{figure}

Figure~\ref{F:azimuthal profile} shows the velocity dispersion as a
function of azimuth about the center of the cluster. As discussed
above, this does not cover the entire $360\degr$ due to the off-center
pointing of HST. Any differences between the radial and tangential
velocity dispersions do not appear significant nor systematic.

\section{Discussion}
\label{discussion}
Our main results are that the central velocity dispersion in \ngc\ is
larger than that measured previously on the basis of radial
velocities, and that there appears to be substantial radial anisotropy
just outside the core of the cluster. The significance of these
results depend on how much credence we give to the velocity dispersion
estimates.

Our analysis of the velocities has been fairly conventional, in that
we have binned the data to derive a velocity dispersion at a number of
distances from the cluster center.  The only unusual feature is the
use of variable uncertainties, but, in the limit where the errors on
the velocities are equal, our method reduces to the quadrature removal
of that error from the standard deviation of the measurements. Our
advantage over \citet{pm93} is that our repeated measurements provide
a good estimate of uncertainties on a star-by-star basis.  The common
use, however, of this technique to explore globular cluster dynamics
has a number of serious conceptual problems.  Here we list and discuss
the difficulties in the standard approach to exploring radial trends
in velocity dispersion.

1) Binning is not, in general, an appropriate way of characterizing
radial change.  There are well-known biases involved in binning,
including unequal distribution of data over the bin, changes in
expected values of a quantity over the bin, the intrinsic trade-off
between resolution (the number of bins) and precision (the number of
data points within a given bin).  Bayesian techniques (e.g.
\citealt{d98}) can be used to get around this problem, but care must
be taken since one is interested in the radial change of a
distribution, rather than of a quantity.

2) In a similar vein, the mass dependence of the velocity distribution
must be taken into account.  Mass segregation is a clear feature of
globular cluster dynamics, and therefore the velocity dispersion
depends on the mass of the objects being measured.  This mass varies
from object to object, and the distribution of masses varies with
radius.  Even objects with very similar observational characteristics
may have quite different masses. Consider, for example, the case of high-mass-ratio,
double-main-sequence, binary stars, which may lie almost on top of the
isochrone for single stars. Since the fraction of binaries doubtless
varies strongly with radius, as in the present case, one may not be
measuring the same mass distribution at different radii even when one
can control for the luminosity distribution.

3) The measurement of the velocity dispersion in a globular cluster is
hampered by the limited number of stars available. This is especially
true in the center of a cluster, where the measurement of the velocity
distribution may be forever limited by the finite number of stars
present. Examination of any of the HST images of the central regions
of globular clusters will quickly illustrate this point. Our proper
motions already cover nearly every star within 11$\arcsec$ of the
center down to well below the main sequence turn off. Since the mass
function in the core of this cluster (like most dense clusters) is
flat, or perhaps even inverted \citep{rb99}, we cannot gain more than
a factor of two by going fainter on the main sequence.  There is a
small vignetted section in the E1994 data, some small regions were
insufficiently observed in the E1999 data, and some stars were lost
due to blending, but the potential gain in total number of stars from
those lost in this way is quite small.  Unless we can detect large
numbers of white dwarfs and neutron stars, the total number of stars
in the core of any given cluster is limited.

4) The shape of the velocity distribution is not in general
understood. One of the most important assumptions in both binned and
Bayesian methods of exploring the velocity distribution in clusters is
that this distribution can be characterized by a single number (the
dispersion) generally taken to be the standard deviation of the
distribution.  Unless the distribution is Gaussian, a dispersion
measured this way is not the same as the second moment of the true
velocity distribution, which could bias further analysis.  But this
assumption of normally distributed velocities must be wrong at some
level.  Even in single-mass isotropic King models the velocity
distribution is {\it not} a Maxwell-Boltzmann distribution, but one
with a cutoff.  In a realistic treatment, it is far from clear what a
measurement of a ``dispersion'' really means.  It is particularly
problematic in small samples (in one bin, for example), where a single
outlier can greatly change the measured value of any moment of a
distribution.

5) Finally, the very concept of a velocity distribution is poorly
defined in a globular cluster.  Defining a ``distribution'' imposes an
assumption of continuity on the population.  In a globular cluster,
and especially in the center, the dynamics may be severely influenced
by non-fluid effects. There are only a finite number of stars, and the
stochastic properties of their particular location in phase space may
be of considerable importance.  This is particularly true given that
the gravitational potential, which gives rise to the velocity
distribution, is created by those very same stars.  The situation is
quite different in the case of the center of a galaxy. In that case,
the black hole is so much more massive than the stars that the
potential is smooth and well-defined, and there are generally enough
stars or gas particles that better observations lead to more
independent measurements in an obvious way. Even in the case of a
galaxy, however, it is not clear that radial velocity measurements
alone are sufficient to independently constrain both the mass-to-light
ratio and the mass of any central black hole \citep{valluri}.

There has been in the past some level of recognition of some of these
problems. None of the solutions to date, however, have been entirely
satisfactory. Alternatives to standard binning have been advocated by
\citet{glowess} and \citet{grotate2} in their LOWESS and ROTATE2
algorithms.  These algorithms attempt to provide a non-parametric
characterization of the velocity-dispersion profile. While the first
explicitly makes a smoothed, non-parametric estimate of the
velocity-dispersion profile, the latter amounts to nothing more than a
maximum-likelihood method applied to a set of overlapping bins of
undisclosed and variable width and population. The derived profiles
are frequently characterized by large, high-frequency
oscillations. The character and location of these oscillations are
easy to replicate using a more standard binning procedure.  As the
uncertainties are taken from bootstrapped samples based on the
inferred profile, the error bands parallel the estimated profile. It
is then hard to decide how much credence to give the apparently
significant fluctuations in the dispersion.

\citet{d98} attempted to use a Bayesian approach to measuring the
dispersion profile both in bins and in various parametric forms.  The
underlying method is similar to the maximum-likelihood method we use
here, but their method of hypothesis comparison is incorrect. What
that attempt at hypothesis comparison does highlight is the wide range
of possible parametrizations and the necessity to objectively choose
the best amongst them. To take some concrete examples, we might find
of use answers to the following sorts of questions: Are our velocities
here better represented by a velocity-dispersion profile consisting of
a single power-law or one with a core? At what level do the radial and
tangential components have profiles that disagree? Ultimately,
however, we would wish to get beyond parametrization to allow the data
to ``speak for themselves''.

\citet{merritt93,merritt96} and \citet{mertrem} have advocated a
series of non-parametric inversion techniques to move from the
observed surface-density profile and projected velocity-dispersion
profile to the more intrinsic density profile, velocity-dispersion
profile, and ultimately the potential and distribution function. These
methods have much to recommend them and a practical application has
been demonstrated in \citet{merritt w cen} in the case of $\omega$
Cen. Merritt's techniques find the minimum of a penalized log
likelihood of the form
\begin{equation}
-\log {\cal L}_p = \sum_{\rm data} \left[{g_p-{ \rm\bf A}g \over \epsilon} \right]^2 + \alpha P\left(g\right),
\end{equation}
where $\rm\bf A$ is a projection operator that brings the desired
function, $g$, into the observable space where we observe $g_p$. The
quantity $\epsilon$ is the measurement error and $P$ is a functional
that assigns large penalties to noisy solutions for $g$.  The degree
of smoothing depends on the value of $\alpha$ and is necessary since
the inversion problem is under-constrained. The function $g$ can be,
for example, the mass-density profile, a velocity or velocity
dispersion, or the distribution function. Until one gets to the
inverted distribution function one is still relying on the assumption
that the velocity distribution is indeed Gaussian in estimating its
dispersion and from this the potential. These techniques give results
that are generally independent of any functional form imposed by
parametrizing assumptions. The resulting functions are those that best
match the data, given the degree of smoothing. They do require large
numbers of velocities and cannot be applied if a more limited quantity
is all that is available. What is lacking is a measure of how convincing a
representation that is.

\citet{gf} use a philosophically similar approach, although quite
differently implemented, to estimate the mass-density profile from the
observed surface-density profile and velocity-dispersion profile
estimated by the LOWESS algorithm \citep{glowess}. They use
spline-smoothed estimates of the profiles which are then numerically
differentiated and integrated to obtain, via Jeans equation, the
mass-density profile. Their method, like Merritt's, recovers the
original model for simulated data. The assumption, though, is that the
velocity ellipsoid is isotropic.

A somewhat more direct approach to measuring the potential of a
globular cluster is seen in \citet{gerssen1,gerssen2} for M15. They
invert the surface-density profile under various assumptions for the
mass-to-light ratio to derive, via Jeans's equation,
velocity-dispersion profiles. They then compare these to the observed
velocities using a maximum-likelihood technique. The large
discrepancies between these profiles and that measured using the
\citet{grotate2} method, despite the supposedly consistently high
likelihood, cannot help but leave us feeling that one or both of the
dispersion estimator and the likelihood argument are incorrect.

Indeed, this work highlights the risk of over-interpreting velocity
data.  \citeauthor{gerssen1} claim to have found evidence for a $3,000
M_\sun$ black hole in the center of M15. This claim is largely
independent of the strength of their claimed fit to the M15
velocity-dispersion profile discussed above. Instead, the evidence in
support of a black hole rests on the velocities of about 12 stars
within 1$\arcsec$ of the cluster center, 8 of these being new HST/STIS
measurements with median error of 5.7 km s$^{-1}$. Even in their
original presentation \citep{gerssen1}, the evidence for a black hole
is, at best, a $1\sigma$ detection. In the final result
\citetext{\citealp{gerssen2} correcting for a scale error in
\citet{dull95} \citep{dull03}}, even the claim for a black hole
becomes ambivalent and dependent on the apparent existence of a
20,000$M_\sun$ black hole in the M31 cluster G1\citep{G1}. But the
claim for G1 is even weaker. The method used for estimating the mass
of the putative black hole there is known to give false minima in
$\chi^2$ in the face of insufficient data even for galaxies
\citep{valluri}. A globular cluster, with its problems of granularity,
stellar sparseness, and small radius of influence for a black hole, is
probably the worst place to apply these techniques.

We do not suggest that we know what is the correct way to proceed
here.  What we would like to do is raise the level of awareness of the
issues involved. What is needed is to identify techniques which work
optimally for a particular quantity and type of data in the face of
the limits discussed above.  Application of these methods must be
tempered by an appreciation of their biases and limitations, and with
proper attention to whether the conclusions drawn are actually
supported by the data or are just artefacts of the method used.

\section{Conclusions}
Our HST proper motions suggest that the velocity dispersion in the
center of \ngc\ is surprisingly large. At 12.5 km s$^{-1}$ it is much
larger than the measured dispersion along the line of sight. While
there is some uncertainty in the distance to \ngc\ it is certainly
known to better than the factor of roughly two which would be required
to bring the two measurements into agreement. Our dispersion indicates
a V-band mass-to-light ratio of 9.  Radial anisotropy appears outside
of about 1.5 core radii. Further observations are required to confirm
the measurements presented here and to extend them to a larger
area. More importantly, the theoretical basis underlying the kinematic
interpretation of stellar velocities in globular cluster needs to be
reconsidered.

\acknowledgements 

We would like to thank J. Anderson for the use of his astrometry code
and him, I.R. King, and S. Shaklan for useful discussions.  The
comments of the anonymous referee and Anderson on the original version
of the appendix prompted us to re-examine our original 34-row
correction. The result is the much improved version presented
below. Drukier and Bailyn were supported by a NASA LTSA grant
NAG5-6404.  van Altena and Girard acknowledge the long-term assistance
of their colleagues on the HST Astrometry Science Team
(W. H. Jefferys, P.I. and G.  F. Benedict, Co-P.I.) who collaborated
in the GTO observations and made it possible for us to obtain the GO
observations that provided the second-epoch observations used in this
investigation.  They also acknowledge financial support from NASA for
the GTO observations and of the STScI for the GO observations.

\appendix\section{34-row correction}
\label{34row}
The so called 34-row error \citep{ak99} originated in the construction
of the WFPC2 CCDs.  It manifests itself as a periodic effect where
every 34th-row of the CCD is a few percent narrower than the
others. In this case we lose uniformity in the $y$-coordinate. Every
34th-row occupies less physical space, but is still assigned the same
unit distance in coordinate space. The net effect is that the observed
$y$ positions of stars deviate systematically from their true
positions in a series of step functions.

\citet{ak99} correct for this using a sawtooth function with period
34.1333 pixels based on measured residuals to a mean position. This
combines a step function with a scale stretch. \citeauthor{ak99} argue
that this is sensible since, if the 34th row is narrower than the
average row, then the other 33 rows are wider. But this is to conflate
the plate scale with the physical source of the defect. And, while 
the correction suggested by \citet{ak99} is correct {\it on
average}, it strikes  us as less than satisfactory. The inter-flaw
spacing, we feel, ought to be an integral number of pixels,
presumably 34. Further, the flawed rows should be identifiable,
allowing for a correction which would be correct in detail, rather
than just on average. To that end, we have reconsidered the nature of
the 34-row flaw and have reexamined the WFPC2 flat fields to test
our reconstruction.

\citet{shaklan} describe the construction of the reticles for the
WFPC2 CCDs in some detail. A 0.5 \micron\ by 0.5 \micron\ e-beam was
scanned in the $y$-direction for 512 \micron\ at a time laying down
the desired pattern. The beam was shifted by 0.5 \micron\ in the
$x$-direction and the scan repeated until the pattern for the full
800-pixel width of the CCD had been laid down on the reticle. The
e-beam was then shifted by 512 \micron\ in $y$ and the next 512
\micron\ region was scanned. This was repeated until the full
800-pixel height of the CCD had been laid down, completing the
pattern. What appears to have happened is that the 512 \micron\
$y$-scans actually occupied 512.5 \micron on the reticle and,
consequently, the CCDs. The shift between vertical scans remained 512
\micron\ so at the edge of each vertical scan a 0.5 \micron\ region
was scanned twice by the e-beam with two consecutive regions of the
CCD pattern. We postulate that the e-beam was magnified by 0.1\% in
the $y$-direction resulting in slightly rectangular pixels, 15
\micron\ wide as planned, but approximately 0.1\% taller. A slight
inclination of the reticle plane with respect to the e-beam could
account for this.

The result is that 512 \micron\ of CCD pattern were replicated onto 512.5
\micron\ of the reticle. At the edge of each scan we find 1 \micron\ of
pattern projects onto 0.5 \micron\ of the physical CCD and mapping from
pattern space to physical position on the CCD is double valued. In
that 0.5 \micron\ region the reverse mapping goes to two parts of the
pattern spaced by $0.5/\delta$ \micron\ where $\delta\sim 1.001$ is the
magnification factor.  This puts a 512 \micron\ periodicity into the
underlying pattern.

In addition, there is a second periodicity, one of 15 \micron,
corresponding to the pixel size. This maps into $15\delta$ \micron\ on
the CCD for the normal pixels.  Every approximately 34 pixels you get
the pixel that contains the doubled-up flaw and, consequently, this
row is narrower by about 0.5 \micron.

The details of which rows are damaged depends on the details of the
relative phasings of the 15 and 512 \micron\ periods and how these map
onto the 800 pixels that define the $y$-dimension of the CCD. We can
certainly see that the 15 \micron\ cycle has phase zero at the edges
of the CCD. It appears, however, that the 512 \micron\ cycle does not
start at the same place.  If you make some assumptions about these
phases, you can easily show that the typical inter-flaw spacing is 34
rows. There is a 15-flaw pattern where, in the 7th and 15th flawed
pixels, the spacing is 35 pixels. There are two cases, where the 34th
pixel is either normal or where both the 34th and 35th pixels are
damaged by a doubled-up region that contains a pixel boundary. Thus,
on average, as noted by \citet{ak99}, the periodicity is
$512/15=34.1\dot 3$ pixels, but this comes from a mode of 34 pixels in
13/15 cases and 35 in the other 2/15. The correction should take into
account the detailed position of the flawed rows rather than only dealing
with the problem on average.

\begin{table}[t]
\caption{\label{T:flaws} Flawed rows}
\begin{tabular}{rr}
\tableline
\tableline
Flaw    &   Row \\
\tableline
   2    &  99		 \\
   3    & 133		 \\
   4    & 167		 \\
   5    & 201		 \\
   6\tablenotemark{d}   & 235-6		\\      
   7\tablenotemark{s}   & 269		 \\				
   7\tablenotemark{p}   & 270		 \\
   8    & 304		 \\
   9    & 338		 \\
  10    & 372		 \\	
  11    & 406		 \\	
  12    & 440		 \\
  13\tablenotemark{p}   & 474		 \\		
  13\tablenotemark{s}   & 475		 \\	
  14\tablenotemark{s}   & 508		 \\	
  14\tablenotemark{p}   & 509		 \\	
  15    & 543		 \\	
  16    & 577		 \\	
  17    & 611		 \\
  18    & 645		 \\
  19    & 679		 \\	
  20    & 713		 \\
  21\tablenotemark{d}   & 747-8		 \\	
  22\tablenotemark{s}   & 781		 \\	
  22\tablenotemark{p}   & 782		 \\	
\tableline
\end{tabular}
\tablenotetext{d}{Doubled flaw. See text.}
\tablenotetext{p}{Primary flaw. See text.}
\tablenotetext{s}{Secondary flaw. See text.}
\end{table}

\begin{figure}[t]
\plotone{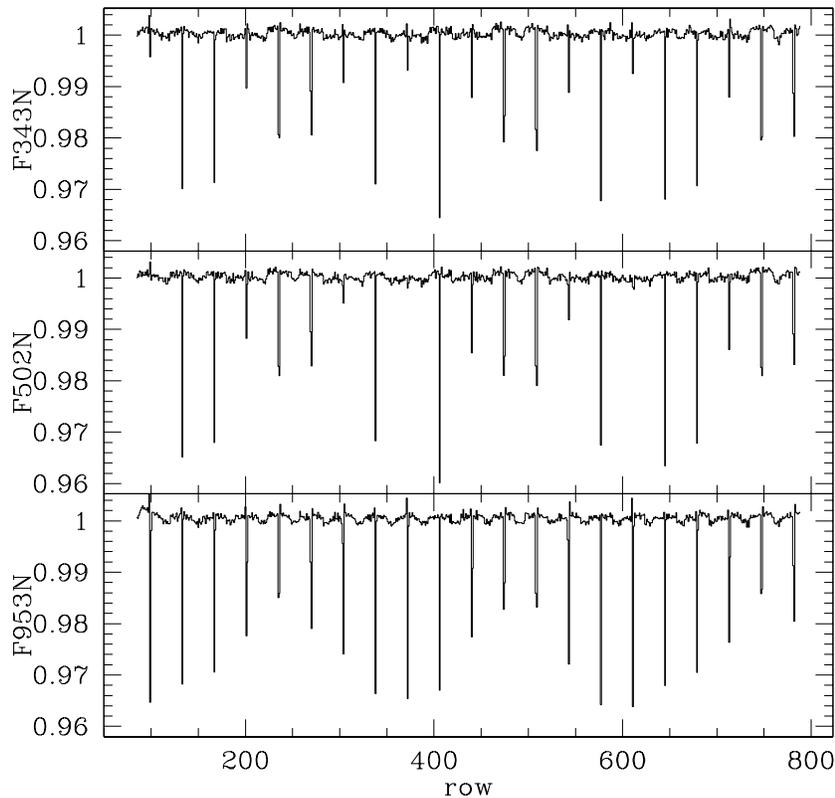}
\caption{The relative response of the WFPC2 CCDs due to
the 34-row flaw pattern. The indicated flats have been divided into
$1\times 25$ median filters of themselves. Note the 15-flaw periodicity
to the pattern and variations in this pattern with wavelength.
\label{F:flat}}
\end{figure}

In order to test our reconstruction, we examined the WFPC2 flat
fields. To do this, we took each flat and divided it into a $1 \times
25$ pixel median filter of itself. We averaged together each row and
all four WFPC2 CCDs since the flaw pattern is identical on
each. Figure~\ref{F:flat} shows the F343N, F502N, and F953N flats
after they have been processed in this manner and the vignetted region
trimmed off. The flawed rows are readily identifiable by their low
intensity.  In this figure, the vertical axis is the relative response
function of the rows due to the 34-row anomaly.  The flawed rows are
listed in Table~\ref{T:flaws}. The flaw number is given by
$\int(15r/512)$, where $r$ is the flawed row. As expected, the modal
inter-flaw spacing is 34 rows. As \citet{ak99} reported in proof from
a private communication from R. Gilliland, there are a number of flaws
which occupy two adjacent rows.  Our model predicts that this would
occur when the 1\micron\ doubling region contains a pixel boundary, but
this should only happen every 15th flaw, i.e.  twice across the CCD,
not the six times seen. We note that the flaw pairs do obey our
expected 15 flaw periodicity as does the relative response at all
wavelengths.

We identify flaws 6 and 21 (rows 235-6 and 747-8) with the expected
situations where the flaw region contains a pixel boundary. The two
rows in each flaw appear to have the same size, roughly half the usual
size.  Flaws 13 and 14, half-way between as expected, also have a
doubled structure, but, in these cases one row is clearly more flawed
than the other. We refer to the row with the deeper signal as primary
and the other as secondary. The primary flaws are in rows 474 and 509,
with the expected 35 row spacing. Other flaws in the vicinity of these
two types of double flaws also show the primary/secondary structure,
particularly flaws 7 and 22. Without understanding the details of the
flawed regions on the level of the CCD electronics, it is impossible
to fully explain this pattern, but it appears to us that the flaw
impacts a region around it to a range of several \micron. If a pixel
boundary falls within the affected region, then the neighboring row
will also appear slightly flawed in the flat fields as seen in
Figure~\ref{F:flat}. The variation in the relative response with
wavelength we also attribute to the electronic structure of the CCD
pixels. Longer-wavelength light penetrates deeper into the CCD, so the
variations with filter are probably telling us something about the
physical properties of the flawed silicon. Note, however, that the
pipeline flat-field frames used here are combined from flat fields in
several filters \citep{flat}. The weakness, or indeed absence, of flaws
2, 10, and 17 at the bluer wavelengths where they are strongest in the
infrared, is particularly interesting. Our model predicts that these
flaws (note the 15-flaw period again) are the most likely to have the
doubled region lying in the center of the pixel. All this confirms the
report by Gilliland in the \citet{ak99} note.  As noted there, these
variations will affect the photometric 34-row correction, but should
not apply to the astrometric correction.

As \citet{shaklan} note, the CCDs, to a high precision, are
square. This is the result of the excess height of the normal pixels
($15\delta \micron$) being compensated for by the narrower, flawed,
rows. The 800 pixels of the CCD are defined by just over 23
repetitions of the 512 \micron\ period, resulting in 23 flawed
rows. The 777 normal rows each exceed their expected size by
$15(\delta - 1)$, so each flawed row is $1-777(\delta-1)/(23\delta)$
tall in units of the $y$-dimension of the pixels. This is 0.966 pixels
for $\delta=1.001$ or 0.967 pixels for $\delta=512.5/512$. We adopt
0.967 for our correction. The difference in $x$ and $y$ scales will be
taken care of in the plate solution.

Our correction is as follows. For each star with centroid not lying in
a flawed row as listed in Table~\ref{T:flaws} we reduce the
$y$-coordinate by 0.033 pixels per flaw with lower $y$ value.  For
stars with centroids lying in a flawed row, the $y$-coordinate is
first rescaled in that pixel to occupy a pixel of size 0.967 rather
than 1.000 before the offset is applied for the previous flawed
rows. For the double-flaws 6 and 21 each row is treated has having
half the flaw size. For the remaining flaws with primary/secondary
structure, only the primary row is considered as being short. A
comparison with the correction of \citet{ak99} shows that, apart from
the scale difference, the maximum difference in the corrected
positions is 0.015 pixels in the flawed rows. In keeping with our
philosophy, the 0.1\% scale difference between $x$ and $y$ is taken
care of in the plate solution.

The data in this paper were corrected with an earlier version of this
correction where a strict 34-pixel spacing was assumed and the
differences in detected positions of stars in offset frames were used
to infer the size of the narrow rows and the phase of the 34-pixel
period with respect to the CCD. The result was a slightly larger
narrow row and slightly different flawed rows at the top and bottom of
the CCD. We have compared the final proper motions with the new
correction and our results are unchanged.


\newpage

\begin{thebibliography}{}

\bibitem[Anderson \& King(1999)]{ak99} Anderson, J. \& King, I. R. 1999, \pasp, 111, 1095
\bibitem[Anderson \& King(2000)]{ak00} Anderson, J. \& King, I. R. 2000, \pasp, 112, 1360
\bibitem[Anderson \& King(2003)]{ak02} Anderson, J. \& King, I. R. 2003, \pasp, 115, 113
\bibitem[Baumgardt \& Makino(2002)]{bm02} Baumgardt, H. \& Makino, J. 2002, \mnras, submitted
\bibitem[Biretta et al.(2001)]{wfpc manual} Biretta, J., et al. 1996, WFPC2 Instrument Handbook, Version 6.0 (Baltimore: STScI)
\bibitem[Cannon et al.(2002)]{cannon} Cannon, R., Da Costa, G., Norris, J., Stanford, L., \& Croke, B. 2002, in ASP Conf. Ser., New Horizons in Globular Cluster Astronomy, ed. G. Pioto, G. Meylan, G. Djorgovski, \& M. Riello, (San Francisco: ASP) in press (astro-ph/0210324)
\bibitem[Carreta et al.(2000)]{carreta}Carretta, E., Gratton, R. G., Clementini, G., \& Fusi Pecci, F. 2000, \apj, 533, 215
\bibitem[Colpi, Possenti, \& Gualandris(2002)]{colpi} Colpi, M., Possenti, A., \& Gualandris, A. 2002, \apj, 570, L85
\bibitem[D'Amico et al.(2002)]{damico1} D'Amico, N., Possenti, A., Fici, L., Manchester, R. N., Lyne, A. G., Camilo, F., \& Sarkissian, J. 2002, \apj, 570, L89
\bibitem[Dinescu et al.(1997)]{dinescu} Dinescu, D. I., Girard, T. M., van Altena, W. F., M\'endez, R. A., \& L\'opez, C .E. 1997, \aj, 114, 1014
\bibitem[Djorgovski \& King(1986)]{dk} Djorgovski, S. \& King, I. R. 1986, \apjl, 305, L61
\bibitem[Drukier et al.(1998)]{d98} Drukier, G. A., Slavin, S. D., Cohn, H. N, Lugger, P., Berrington, R. C., Murphy, B. W., \& Seitzer, P. O. 1998, \aj, 115, 708
\bibitem[Dubath, Meylan, \& Mayor(1997)]{dmm97} Dubath, P., Meylan, G. \& Mayor, M. 1997, \aap, 324, 505 
\bibitem[Dull et al.(1997)]{dull95} Dull, J. D., Cohn, H. N., Lugger, P. M., Murphy, B. W., Seitzer, P. O., Callanan, P. J., Rutten, R. G. M., \& Charles, P. A. 1997, \apj, 481, 267
\bibitem[Dull et al.(2003)]{dull03} Dull, J. D., Cohn, H. N., Lugger, P. M., Murphy, B. W., Seitzer, P. O., Callanan, P. J., Rutten, R. G. M., \& Charles, P. A. 2003, \apj, in press 
\bibitem[Gao et al.(1991)]{gao} Gao, B., Goodman, J., Cohn, H., \& Murphy, B. 1991, \apj, 370, 567
\bibitem[Gebhardt \& Fischer(1995)]{gf} Gebhardt, K. \& Fischer, P. 1995, \aj, 109, 209
\bibitem[Gebhardt et al.(1994)]{glowess} Gebhardt, K., Pryor, C., Williams, T. B., \& Hesser, J. E. 1994, \aj, 107, 2067
\bibitem[Gebhardt et al.(2000)]{grotate2} Gebhardt, K., Pryor, C., O'Connell, R. D., Williams, T. B., \& Hesser, J. E. 2000, \aj, 119, 1268
\bibitem[Gebhardt, Rich, \& Ho(2002)]{G1} Gebhardt, K., Rich, R. M., \& Ho, L. C. 2002, \apj, 578, L41
\bibitem[Gerssen et al.(2002)]{gerssen1} Gerssen, J., van der Marel, R. P., Gebhardt, K., Guhathakurta, P., Peterson, R. C., \& Pryor, C. 2002, \aj, 124, 3270
\bibitem[Gerssen et al.(2003)]{gerssen2} Gerssen, J., van der Marel, R. P., Gebhardt, K., Guhathakurta, P., Peterson, R. C., \& Pryor, C. 2003, \aj, 125, 376
\bibitem[Gratton et al.(2001)]{gratton} Gratton, R. G., et al. 2001, \aap, 369, 87
\bibitem[Grindlay et al.(2001)]{grindlay} Grindlay, J. E., Heinke, G., Edmonds, P. D., \& Murray, S. S. 2001, Science, 292, 2290
\bibitem[Harbeck, Smith, \& Grebel(2003)]{harbeck} Harbeck, D., Smith, G. H., \& Grebel, E. K. 2003, \aj, 125, 197
\bibitem[Koekemoer, Biretta, \& Mack(2002)]{flat} Koekemoer, A.M., Biretta, J. \& Mack, J. 2002, WFPC2 Instrument Science Report 2002-02.
\bibitem[McLaughlin(2000)]{mclau} McLaughlin, D. E. 2000, \apj, 539, 618
\bibitem[Merritt(1993)]{merritt93} Merritt, D. 1993, in Structure, Dynamics and Chemical Evolution of Elliptical Galaxies, ed. I. J. Danziger, W. W. Zeilinger, \&  K. Kj\"ar, (Munich: ESO), 275
\bibitem[Merritt(1996)]{merritt96} Merritt, D. 1996, \aj, 112, 1085
\bibitem[Merritt, Meylan, \& Mayor(1997)]{merritt w cen} Merritt, D., Meylan, G., \& Mayor, M. 1997, \aj, 114, 1074 
\bibitem[Merritt \& Tremblay(1994)]{mertrem} Merritt, D. \& Tremblay, B. 1994, \aj, 108, 514
\bibitem[Momany et al.(2002)]{mom02} Momany, Y., Piotto, G., Recto-Blanco, A., Bedin, L. R., Cassisi, S., \& Bono, G. 2002, \apj, 576, L65
\bibitem[Nemec \& Nemec(1993)]{nemec2} Nemec, J. M. \& Nemec, A. F. L. 1993, \aj, 105, 1455
\bibitem[Norris et al.(1981)]{bimodal}Norris, J. E., Cottrell, P. L., Freeman, K. C., DaCosta, G. S. 1981,\apj, 244, 205
\bibitem[Pooley et al.(2002)]{pooley} Pooley, D. et al. 2002, \apj, 569, 405
\bibitem[Pryor \& Meylan(1993)]{pm93} Pryor, C. \& Meylan, G. 1993, in ASP Conf. Ser. 50, Structure and Dynamics of Globular Clusters, ed. S.G. Djorgovski \& G. Meylan, (San Francisco: ASP), 357
\bibitem[Renzini et al.(1996)]{renzini} Renzini, A., Bragaglia, A., Ferraro, F. R., Gilmozzi, R., Ortolani, S., Holberg, J. B., Liebert, J., Wesemael, F., \& Bohlin, R. C. 1996, \apj, 465, L23
\bibitem[Rubenstein \& Bailyn(1997)]{rb97} Rubenstein, E. P. \& Bailyn, C. D. 1997, \apj, 474, 701
\bibitem[Rubenstein \& Bailyn(1999)]{rb99} Rubenstein, E. P. \& Bailyn, C. D. 1999, \apj, 513, L33
\bibitem[Shaklan et al.(1995)]{shaklan} Shaklan, S., Sharman, M. C. \& Pravdo, S. H. 1995, Applied Optics, 34, 6672
\bibitem[Shawl \& White(1986)]{sw86} Shawl S. J. \& White R. E. 1986, \aj, 91, 312
\bibitem[Trager, King, \& Djorgovski(1995)]{tkd} Trager, S. C., King, I. R., \& Djorgovski, S. 1995, \aj, 109, 218
\bibitem[Valluri, Merritt, \& Emsellem(2002)]{valluri} Valluri, M., Merritt, D., \& Emsellem, E. 2002,  \apj, submitted (astro-ph/0210379)

\end{thebibliography}
\end{document}